\documentclass[prl,aps,preprint]{revtex4-1}

\def\r{{\vec r}}

  \usepackage{graphicx}
  \usepackage{dcolumn}
  \usepackage{bm}
\begin{document}

\title{Natural Orbital Functional Theory and 
Pairing Correlation Effects in Electron Momentum Density}

\author{B. Barbiellini}

\affiliation{Department of Physics, 
Northeastern University, 
Boston, MA 02115 USA}

%
\begin{abstract}
Occupation numbers of natural orbitals
capture the physics of strong electron 
correlations in momentum space.
A Natural Orbital Density Functional Theory
based on the antisymmetrized geminal product
provides these occupation numbers and the corresponding
electron momentum density.
A practical implementation of this theory
approximates the natural orbitals by the 
Kohn-Sham orbitals and uses a 
mean-field approach to estimate pairing 
amplitudes leading to corrections for 
the independent
particle model.
The method is applied to weakly 
doped La$_2$CuO$_4$.

\end{abstract}


\maketitle

\vskip 1cm

A key characteristic of an interacting electron system 
is the electron momentum density (EMD). For metallic systems 
one can also define the Fermi Surface (FS) 
as the break in the EMD whose presence reveals 
the existence of quasi-particles and 
the validity of the 
Landau-Fermi liquid theory \cite{khomskii}.
FS studies are particular needed in the field
of high temperature superconductivity.
Figure 1 shows the calculated FS of the 
well known HgBa$_2$CuO$_4$ \cite{bba1993}
while Figure 2 illustrates the evolution 
of the FS topology with doping 
of a less known compound studied 
by Jarlborg {\em et al.} \cite{jarlborg2012}.
One can notice in Fig.~2 a topological
transition (also called Lifshitz transition),
which does not involve any 
symmetry breaking \cite{lifshitz1960}.

Positron annihilation has been successful for 
the determination of the 
FS in many metallic systems,
but similar studies of the copper oxide high 
temperature superconductors 
have met difficulties since positrons do not 
probe well the FS contribution 
of the Cu-O planes \cite{icpa16}. 
Another direct probe of FS is the Angular 
Resolved Photo-Emission Spectroscopy (ARPES)
\cite{ARPES}.  However, a concern with ARPES is that most 
of the information of the interacting electron liquid 
is based on measurements from a surface sensitive        
technique that can be applied only to a 
limited number of materials that cleave
such as Bi$_2$Sr$_2$CaCu$_2$O$_{8-\delta}$.  
Thus, a risk is that experimental artifacts may be 
interpreted as fundamental physics. 

At low doping, the FS signal from ARPES breaks up into 
Fermi arcs \cite{norman1998}, 
which could be part of closed hole pockets \cite{yang2008,yang2011}.
The formation of small Fermi pockets in other underdoped cuprates 
also emerges from quantum oscillation (QO) measurements 
in high magnetic fields
\cite{leboeuf2007,2,3a,3b,sebastian2010}.
These FS pieces seen by QOs could be in fact produced 
by FS reconstructions when some symmetry is broken \cite{sudip}. 
According the theory by Lifshitz and Kosevich \cite{LK,LK2},
a period of QO is linked to an extreme cross section of the FS. 
Nevertheless, QOs in the layered and quasi-two-dimensional (2D) 
conductors may deviate from the LK theory developed 
for 3D conventional metals 
\cite{LKM,hartnoll}.

\begin{figure}
\begin{center}
\includegraphics[width=8cm,height=8cm]{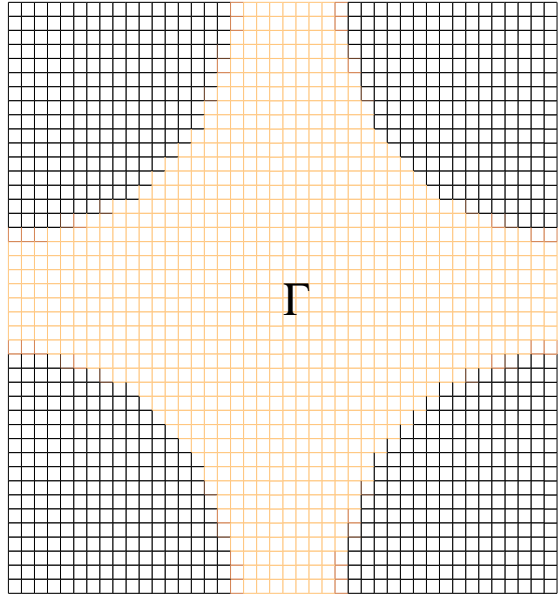}
\end{center}
\caption{The FS of HgBa$_2$CuO$_4$ is shown in the first Brillouin zone.
It separates
the occupied states (yellow grid) from the 
unoccupied states (black grid).} 
\label{fighbco}
\end{figure}

\begin{figure}
\begin{center}
\includegraphics[width=8cm,height=8cm]{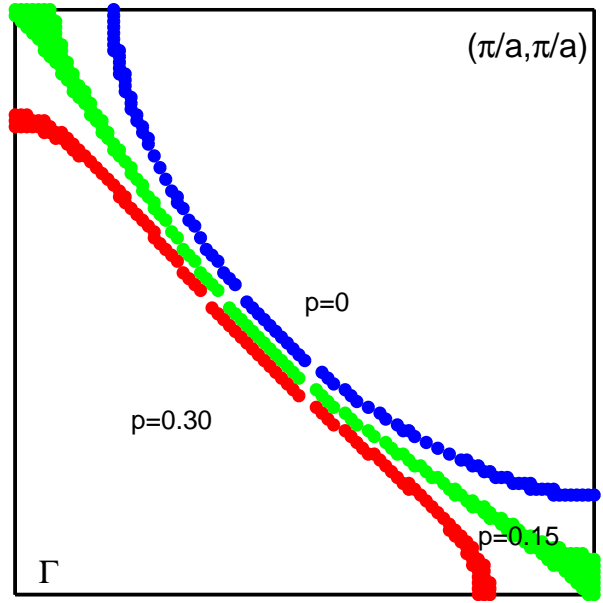}
\end{center}
\caption{ Evolution of the Fermi surface 
in the $k_z = 0$-plane in Ba$_2$CuO$_3$ 
(where one layer of apical oxygen is missing)
as a function of the rigid-band doping for 0, 0.15 and 
0.30 holes per unit cell. The FS evolution is almost identical 
in La$_2$CuO$_4$. Only 1/4 of the first Brillouin zone is shown.
The momentum units are $1/a$, where $a$ is the lattice constant.} 
\label{figFSD}
\end{figure}

Inelastic x-ray scattering 
\cite{ray,piero} in the deeply inelastic limit, 
can help to clarify the nature of the FS in
copper oxide high temperature superconductors
since the corresponding Compton scattering cross-section 
is well-known to become proportional to the ground 
state EMD \cite{kaplan}.
A Compton scattering study in 
single crystals of La$_{2-x}$Sr$_{x}$CuO$_{4}$ has directly 
imaged in momentum space 
the character of holes doped into this material \cite{science}.
However, improvements in the momentum resolution are still 
needed to bring Compton scattering into the fold of 
mainstream probes for the cuprates FS. 
A recent Compton study of overdoped La$_{2-x}$Sr$_2$CuO$_4$ 
\cite{wael} shows the difficulty of         
extracting details of the FS with 
the present momentum resolution of about $0.15$ a.u.
Higher momentum resolution can also allow 
the study of the FS smearing due to 
the superconducting energy gap opening \cite{peter} 
and to the breakdown of the Landau-Fermi liquid picture 
\cite{bba1}.
Surprisingly, Compton scattering experiments 
even on a simpler material
such Li indicate that the EMD of the ground state is not well
described by the conventional Landau-Fermi liquid framework
since the size of the discontinuity $Z$ at the FS
seems to be anomalously small \cite{li1,li2,li3,li4}.
Such deviations from the standard metallic picture can
be ascribed to the possible existence of 
significant pairing correlations in
the ground state \cite{bba1,momentum,dan}. 
The notion of stabilizing the metallic state through the
resonant valence bond (RVB) state dates back
to the early works of Pauling, who first applied this model to the Li ground
state \cite{paulingLi4}. Anderson then proposed the RVB wave function as
a ground state for the high temperature superconducting materials
\cite{anderson}, showing that this hypothesis is able of describing many
aspects of the phase diagram of the cuprates \cite{vanilla}.

This paper shows how pairing correlation effects modify
the occupation of the natural orbitals \cite{lowdin,goscinki}, 
which are used to calculate
the electron momentum density \cite{momentum}
via a simple generalization
of independent particle model (IPM).
In the IPM, states are either occupied or empty \cite{kohn96}.
 
The occupation numbers
(corresponding to natural orbitals 
in Bloch states \cite{calais1988,lcw})
are calculated through a variational scheme
based on the Antisymmetrized Geminal Product (AGP) many-body wave function
\cite{bba1,momentum,blatt1,blatt,bratoz,coleman,goscinski,marel}.
The AGP total energy functional is given by \cite{momentum,blatt,goscinski}
\begin{equation}
E_{AGP}=E_{HF} +  E_{BCS} + O(1/N)~,
\label{toten}
\end{equation}
where $E_{HF}$ is the Hartree Fock energy functional,
$E_{BCS}$ is a BCS-type functional and $N$ 
is the number of electrons in the system.
Since the Coulomb interaction contained in $E_{BCS}$ is 
repulsive, energy can be gained only through the 
exchange part $E_{HF}$ of the Hartree-Fock functional. 
However, energy can also be gained 
through the term $E_{BCS}$  
by the introduction interactions with phonons \cite{weger}.
Several authors 
\cite{gu1,gu2,gabor,BB02,GPB05,HLA05,SDLG08,piris2013} 
have considered similar functionals of natural orbitals.
Nevertheless, some of these functionals violate the $N$-representability 
\cite{momentum} and can become over-correlated.
This problem is avoided here because the 
$E_{AGP}$ is $N$-representable by construction.

To efficiently extract occupation numbers 
in a correlated electron gas, 
we start by approximating 
the natural orbitals by the Kohn-Sham orbitals \cite{ks}.  
For the sake of simplicity, we suppose that
the eigenvalues are described by a single energy band
denoted by ${\cal E}_{\vec k}$ with $\mu$ 
defining the chemical potential.
The result for the AGP energy functional  
\cite{momentum} minimization gives the
occupation numbers
\begin{equation}
n_{\vec k}= {1\over2}\left( 1 -
{({\cal E}_{\vec k} - \mu)
\over E_{\vec k}}\right),
\label{renocc}
\end{equation}
where $E_{\vec k}$ is given by
\begin{equation}
E_{\vec k} = \sqrt{({\cal E}_{\vec k} - \mu)^2 + 
|\Delta_{\vec k}|^2}~.
\end{equation}
Two self-consistent equations are also
involved, one giving $\Delta_{\vec k}$ \cite{hpa}
\begin{equation}
\Delta_{\vec k}={1\over N}\sum_{\vec k^\prime} \frac{J_{\vec k\vec k^\prime}
\Delta_{\vec k^\prime}}{2E_{\vec k}},
\label{gap}
\end{equation}
and the other determining the chemical potential $\mu$ 
\begin{equation}
N=\sum_{\vec k} n_{\vec k}.
\label{mu}
\end{equation}
Following Ref.~\cite{momentum},
one can assume that
$J_{\vec k\vec k^\prime}$ 
is mostly given by an exchange integral.
Thus, an approximation for $J_{\vec k\vec k^\prime}$ 
is given by \cite{momentum,janak}
\begin{equation}
J_{\vec k\vec k^\prime}=\delta_{\vec k\vec k^\prime} I_{\vec k}~,
\end{equation}
with 
\begin{equation}
I_{\vec k}=\frac{1}{3} \int d^3\r 
|\psi_{\vec k}(\r)|^4 \frac{v_x( \r)}{n(\r)}~,
\label{eq:janak}
\end{equation}
where $v_x( \r)=2/\pi(3\pi^2n(\r)^{1/3})$ is the Kohn-Sham
exchange potential \cite{ks} and $n(\r)$ is the electron
density. 
By inserting this approximation in Eq.~\ref{gap}, one obtains
\begin{equation}
\Delta_{\vec k}^2=\frac{I_{\vec k}^2-4({\cal E}_{\vec k} - \mu)^2}{4}.
\end{equation}
Therefore $\Delta_{\vec k}$ is different of zero only if 
$I_{\vec k} > 2({\cal E}_{\vec k} - \mu)$.

The calculation of the occupation numbers for La$_{2-x}$Sr$_{x}$CuO$_{4}$
shown in Fig.~3  has been performed within an efficient linear muffin-tin 
orbital (LMTO) band structure method \cite{lmto}. 
In this case, 
the average $I$ is about $1.67$ eV.
Therefore, at the Fermi energy $\Delta \sim I/2= 0.83$ eV.
The momentum smearing produced by $\Delta$ is given by
\begin{equation}
\delta k = \frac{\Delta}{v_F},
\end{equation}
where $v_F$ is the Fermi velocity. 
By taking $\hbar v_F\sim \pi/a$ (where
$a=7.16$ a.u. is the lattice constant) we find 
$\delta k \sim 0.07$ a.u. This momentum smearing is slightly below
the current experimental momentum resolution of $0.15$ a.u. 
available in Compton scattering experiments \cite{wael}.
A similar $\delta k$ can be produced by 
the antiferromagnetic order \cite{FP1989} 
when $x \rightarrow 0$,
but the emergence of ferromagnetic fluctuations
for $x \sim 0.25$ 
leads to the destruction of both RVB correlations 
and of the AF order in the over-doped regime \cite{bba2008}.

\begin{figure}
\begin{center}
\includegraphics[width=8cm,height=8cm]{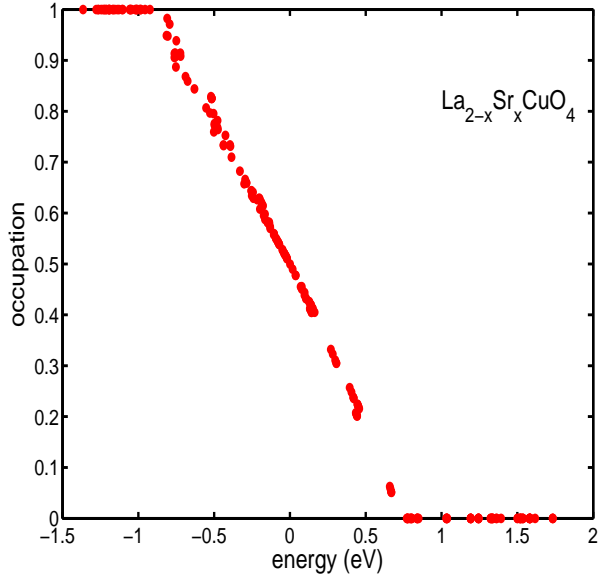}
\end{center}
\caption{Occupation number for La$_{2-x}$Sr$_{x}$CuO$_{4}$
in the limit $x \rightarrow 0$ as a function of the 
Kohn-Sham eigenvalues.
The Fermi level is at 0.} 
\label{figOCC}
\end{figure}

In conclusion, the AGP method has been used to study the occupation 
numbers $n_{\vec k}$ of natural orbitals in Bloch states of crystals.
Values of $n_{\vec k}$ can be extracted from EMD experiments
\cite{lcw} and compared to the present model.
Strong modification of the occupation numbers due to pairing 
correlations effects are
predicted for La$_{2-x}$Sr$_{x}$CuO$_{4}$ with $x \rightarrow 0$
in an energy window of $0.83$ eV around the Fermi energy.
This effect produces a smearing of the occupation in momentum space  
given by $\delta k \sim 0.07$ a.u.

\begin{acknowledgments}
This work is supported by the US Department of Energy, Office of Science, 
Basic Energy Sciences Contract No. DE-FG02-07ER46352. It has also benefited 
from Northeastern University's Advanced Scientific Computation Center (ASCC), 
theory support at the Advanced Light Source, Berkeley, and the allocation of 
computer time at NERSC through Grant No. DE-AC02-05CH11231.
\end{acknowledgments}

\end{document}